\PassOptionsToPackage{unicode}{hyperref}
\PassOptionsToPackage{hyphens}{url}
\PassOptionsToPackage{dvipsnames,svgnames,x11names}{xcolor}
\documentclass[
]{article}
\usepackage{amsmath,amssymb}
\usepackage{lmodern}
\usepackage{iftex}
\ifPDFTeX
  \usepackage[T1]{fontenc}
  \usepackage[utf8]{inputenc}
  \usepackage{textcomp} % provide euro and other symbols
\else % if luatex or xetex
  \usepackage{unicode-math}
  \defaultfontfeatures{Scale=MatchLowercase}
  \defaultfontfeatures[\rmfamily]{Ligatures=TeX,Scale=1}
\fi
\IfFileExists{upquote.sty}{\usepackage{upquote}}{}
\IfFileExists{microtype.sty}{% use microtype if available
  \usepackage[]{microtype}
  \UseMicrotypeSet[protrusion]{basicmath} % disable protrusion for tt fonts
}{}
\makeatletter
\@ifundefined{KOMAClassName}{% if non-KOMA class
  \IfFileExists{parskip.sty}{%
    \usepackage{parskip}
  }{% else
    \setlength{\parindent}{0pt}
    \setlength{\parskip}{6pt plus 2pt minus 1pt}}
}{% if KOMA class
  \KOMAoptions{parskip=half}}
\makeatother
\usepackage{xcolor}
\usepackage{graphicx}
\makeatletter
\def\maxwidth{\ifdim\Gin@nat@width>\linewidth\linewidth\else\Gin@nat@width\fi}
\def\maxheight{\ifdim\Gin@nat@height>\textheight\textheight\else\Gin@nat@height\fi}
\makeatother
\setkeys{Gin}{width=\maxwidth,height=\maxheight,keepaspectratio}
\makeatletter
\def\fps@figure{htbp}
\makeatother
\providecommand{\tightlist}{%
  \setlength{\itemsep}{0pt}\setlength{\parskip}{0pt}}
\newlength{\cslhangindent}
\newlength{\csllabelwidth}
\newlength{\cslentryspacingunit} % times entry-spacing
\newenvironment{CSLReferences}[2] % #1 hanging-ident, #2 entry spacing
 {% don't indent paragraphs
  \setlength{\parindent}{0pt}
  \ifodd #1
  \let\oldpar\par
  \def\par{\hangindent=\cslhangindent\oldpar}
  \fi
  \setlength{\parskip}{#2\cslentryspacingunit}
 }%
 {}
\usepackage{calc}

\ifLuaTeX
\usepackage[bidi=basic]{babel}
\else
\usepackage[bidi=default]{babel}
\fi
\babelprovide[main,import]{american}
\def\languageshorthands#1{}
\ifLuaTeX
  \usepackage{selnolig}  % disable illegal ligatures
\fi
\IfFileExists{bookmark.sty}{\usepackage{bookmark}}{\usepackage{hyperref}}
\IfFileExists{xurl.sty}{\usepackage{xurl}}{} % add URL line breaks if available
\hypersetup{
  pdftitle={superblockify: A Python Package for Automated Generation,
Visualization, and Analysis of Potential Superblocks in Cities},
  pdfauthor={Carlson M. Büth, Anastassia Vybornova, Michael Szell},
  pdflang={en-US},
  colorlinks=true,
  linkcolor={Maroon},
  filecolor={Maroon},
  citecolor={Blue},
  urlcolor={Blue},
  pdfcreator={LaTeX via pandoc}}

\title{\texttt{superblockify}: A Python Package for Automated
Generation, Visualization, and Analysis of Potential Superblocks in
Cities}

\usepackage[affil-it]{authblk}
\usepackage{orcidlink}
\author[1,2%
  ]{Carlson M. Büth%
    \,\orcidlink{0000-0003-2298-8438}\,%
    }
\author[1%
  ]{Anastassia Vybornova%
    \,\orcidlink{0000-0001-6915-2561}\,%
    }
\author[1,3,4%
  ]{Michael Szell%
    \,\orcidlink{0000-0003-3022-2483}\,%
    }

\affil[1]{NEtwoRks, Data, and Society (NERDS), Computer Science
Department, IT University of Copenhagen, 2300 Copenhagen, Denmark}
\affil[2]{Institute for Cross-Disciplinary Physics and Complex Systems
(IFISC), University of the Balearic Islands (UIB) and Spanish National
Research Council (CSIC), 07122 Palma de Mallorca, Spain}
\affil[3]{ISI Foundation, 10126 Turin, Italy}
\affil[4]{Complexity Science Hub Vienna, 1080 Vienna, Austria}
\date{23 April 2024}

\begin{document}
\maketitle

\hypertarget{summary}{%
\section{Summary}\label{summary}}

\texttt{superblockify} is a Python package for partitioning an urban
street network into Superblock-like neighborhoods and for visualizing
and analyzing the partition results. A Superblock is a set of adjacent
urban blocks where vehicular through traffic is prevented or pacified,
giving priority to people walking and cycling
(\protect\hyperlink{ref-nieuwenhuijsen_superblock_2024}{Nieuwenhuijsen
et al., 2024}). The Superblock blueprints and descriptive statistics
generated by \texttt{superblockify} can be used by urban planners as a
first step in a data-driven planning pipeline, or by urban data
scientists as an efficient computational method to evaluate Superblock
partitions. The software is licensed under AGPLv3 and is available at
\url{https://superblockify.city}.

\hypertarget{statement-of-need}{%
\section{Statement of need}\label{statement-of-need}}

The Superblock model is an urban planning intervention with massive
public health benefits that creates more liveable and sustainable cities
(\protect\hyperlink{ref-laverty2021low}{Laverty et al., 2021};
\protect\hyperlink{ref-mueller2020}{Mueller et al., 2020};
\protect\hyperlink{ref-world2022walking}{WHO, 2022}). Superblocks form
human-centric neighborhoods with reduced vehicular traffic. They are
safer, quieter, and more environmentally friendly
(\protect\hyperlink{ref-agenciadecologiaurbanadebarcelona2021}{Agència
d'Ecologia Urbana de Barcelona et al., 2021};
\protect\hyperlink{ref-martin2021}{Martin, 2021};
\protect\hyperlink{ref-mueller2020}{Mueller et al., 2020}) than
car-centric urban landscapes which fully expose citizens to car harm
(\protect\hyperlink{ref-miner2024car}{Miner et al., 2024}). The
scientific study of Superblocks has expanded quickly in recent years,
summarized in a review by Nieuwenhuijsen et al.
(\protect\hyperlink{ref-nieuwenhuijsen_superblock_2024}{2024}). The
planning and implementation of Superblocks is an intricate process,
requiring extensive stakeholder involvement and careful consideration of
trade-offs (\protect\hyperlink{ref-nieuwenhuijsen2019}{Nieuwenhuijsen et
al., 2019}; \protect\hyperlink{ref-stadtwien2021}{Stadt Wien, 2021};
\protect\hyperlink{ref-transportforlondon2020}{Transport for London,
2020}). New computational tools and data sets, such as the
\texttt{osmnx} Python library
(\protect\hyperlink{ref-boeing2017}{Boeing, 2017}) and OpenStreetMap
(\protect\hyperlink{ref-openstreetmapcontributors2023}{OpenStreetMap
contributors, 2023}), provide the opportunity to simplify this process
by allowing to easily analyze and visualize urban street networks
computationally. Recent quantitative studies on Superblocks have seized
this opportunity with different focuses, such as potential Superblock
detection via network flow on the abstract level
(\protect\hyperlink{ref-eggimann_potential_2022}{Eggimann, 2022a}) or in
the local context of Vienna
(\protect\hyperlink{ref-frey2020potenziale}{Frey et al., 2020});
development of interactive micro-level planning tools
(\protect\hyperlink{ref-abstreet}{Carlino et al., 2024};
\protect\hyperlink{ref-tuneourblock}{\emph{TuneOurBlock}, 2024}); green
space (\protect\hyperlink{ref-eggimann_expanding_2022}{Eggimann,
2022b}), social factors (\protect\hyperlink{ref-yan_redefining_2023}{Yan
\& Dennett, 2023}), health benefit modeling
(\protect\hyperlink{ref-li_modeling_2023}{Li \& Wilson, 2023}), or an
algorithmic taxonomy of designs
(\protect\hyperlink{ref-feng_algorithmic_2022}{Feng \& Peponis, 2022}).
However, to our knowledge, none of these emerging research efforts have
led to an open, general-use, extendable software package for Superblock
delineation, visualization, and analysis. \texttt{superblockify} fills
this gap.

The software offers benefits for at least two use cases. First, for
urban planning, it provides a quick way to generate Superblock
blueprints for a city, together with descriptive statistics informing
the planning process. These blueprints can serve as a vision or first
draft for potential future city development. In a planning pipeline,
\texttt{superblockify} stands at the beginning, broadly delineating the
potential areas of study first. Then, exported Superblocks can feed into
an open geographic information system like QGIS
(\protect\hyperlink{ref-qgis}{QGIS Development Team, 2024}) or into
tools like A/B Street (\protect\hyperlink{ref-abstreet}{Carlino et al.,
2024}) or TuneOurBlock
(\protect\hyperlink{ref-tuneourblock}{\emph{TuneOurBlock}, 2024}) that
allow finetuned modifications or traffic simulations. This quick
feedback can reduce the time and resources required to manually plan
Superblocks, which in turn can accelerate sustainable urban development.
Second, \texttt{superblockify} enables researchers to conduct
large-scale studies across multiple cities or regions, providing
valuable insights into the potential impacts of Superblocks at a broader
scale. In both cases, \texttt{superblockify} can help to identify best
practices, algorithmic approaches, and strategies for Superblock
implementation.

The software has served in a preliminary analysis of potential
Superblocks in 180 cities worldwide
(\protect\hyperlink{ref-buth2023master}{Büth, 2023}) and will be used in
subsequent studies within the EU Horizon Project JUST STREETS
(\url{https://just-streets.eu}). With increased urbanization, impacts of
climate change, and focus on reducing car-dependence
(\protect\hyperlink{ref-mattioli_political_2020}{Mattioli et al., 2020};
\protect\hyperlink{ref-ritchie2018}{Ritchie \& Roser, 2018};
\protect\hyperlink{ref-satterthwaite2009}{Satterthwaite, 2009}), the
need for sustainable urban planning tools like \texttt{superblockify}
will only increase
(\protect\hyperlink{ref-nieuwenhuijsen_superblock_2024}{Nieuwenhuijsen
et al., 2024}).

\hypertarget{features}{%
\section{Features}\label{features}}

\texttt{superblockify} has three main features: Data access and
partitioning, Visualization, and Analysis.

\hypertarget{data-access-and-partitioning}{%
\subsection{Data access and
partitioning}\label{data-access-and-partitioning}}

\texttt{superblockify} leverages OpenStreetMap data
(\protect\hyperlink{ref-openstreetmapcontributors2023}{OpenStreetMap
contributors, 2023}) and population data GHS-POP R2023A
(\protect\hyperlink{ref-pesaresi2023}{Pesaresi \& Politis, 2023}). From
a user-given search query, e.g., a city name, \texttt{superblockify}
retrieves the street network data of a city, the necessary GHS-POP
tile(s), and distributes the population data onto a tesselation of the
street network.

After the street network and optional metadata are loaded in, the
package partitions the street network into Superblocks. In its current
version 1.0.0, \texttt{superblockify} comes with two partitioners:

\begin{enumerate}
\def\labelenumi{\arabic{enumi}.}
\tightlist
\item
  The residential approach uses the given residential street tag to
  decompose the street network into Superblocks.
\item
  The betweenness approach uses the streets with high betweenness
  centrality for the decomposition.
\end{enumerate}

The resulting Superblocks can be exported in GeoPackage (\texttt{.gpkg})
format for further use.

\hypertarget{visualization}{%
\subsection{Visualization}\label{visualization}}

\begin{figure}
\centering
\includegraphics{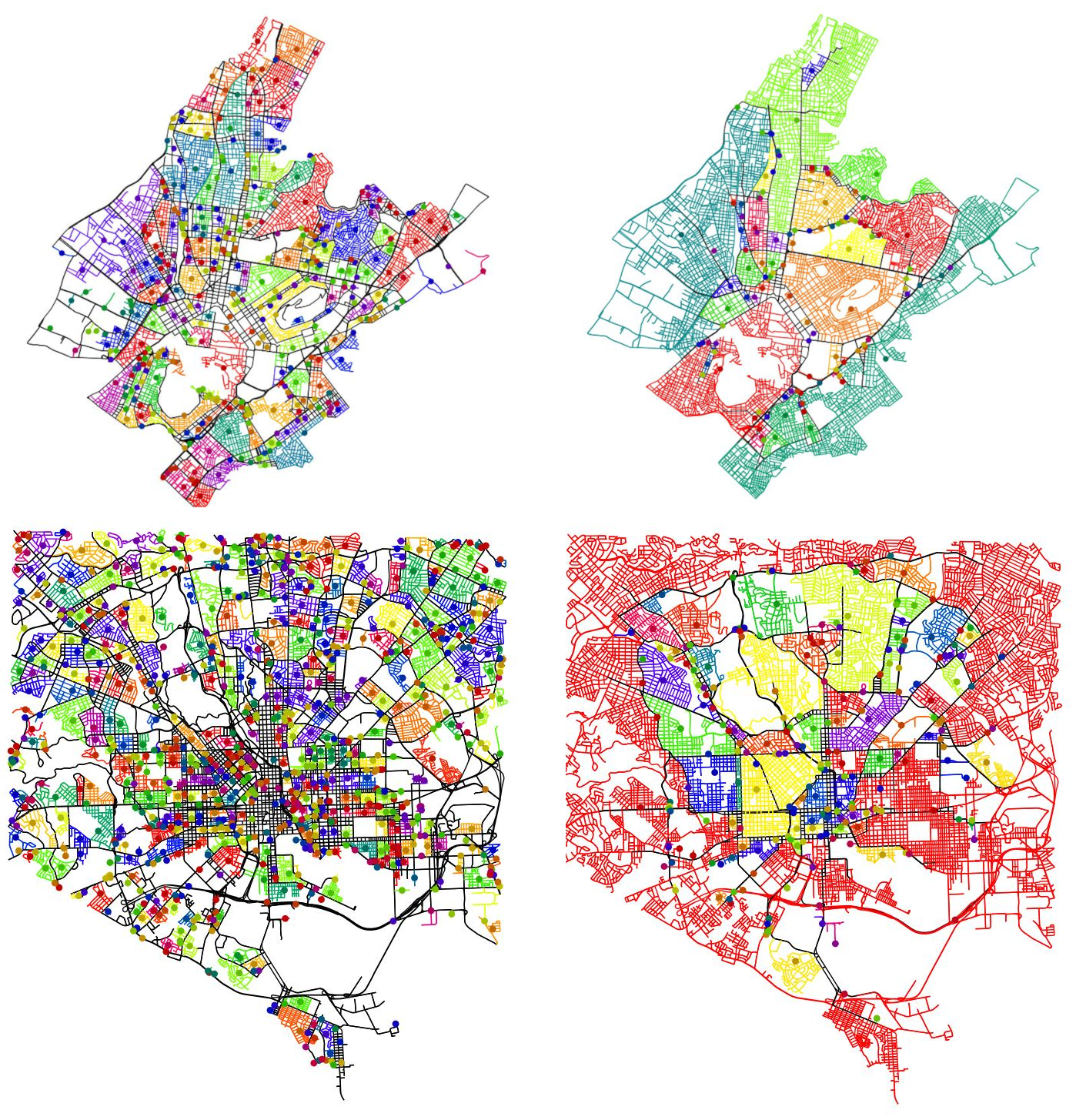}
\caption{Automated generation of Superblocks. Athens (top row) and
Baltimore (bottom row) Superblocks generated using the residential
partitioner (left column) and the betweenness partitioner (right
column). The streets of each Superblock are colored, the rest of the
streets are black. Colored nodes denote representative nodes within each
Superblock for easier visual recognition. Map data from OpenStreetMap.
\label{fig:combined_graphs}}
\end{figure}

After the partitioning, factors relevant for analysis and planning of
Superblocks can be calculated and visualized, e.g., area, population,
population density, or demand change by betweenness centrality. Example
Superblock configurations for two cities are shown in Fig.
\ref{fig:combined_graphs}.

\hypertarget{analysis}{%
\subsection{Analysis}\label{analysis}}

For analysis, the package calculates various graph metrics of the street
network, such as global efficiency
(\protect\hyperlink{ref-latora2001}{Latora \& Marchiori, 2001}),
directness (\protect\hyperlink{ref-szell2022}{Szell et al., 2022}),
betweenness centrality (\protect\hyperlink{ref-brandes2008}{Brandes,
2008}), spatial clustering and anisotropy of high betweenness centrality
nodes (\protect\hyperlink{ref-kirkley2018}{Kirkley et al., 2018}),
street orientation-order (\protect\hyperlink{ref-boeing2019}{Boeing,
2019b}), or average circuity
(\protect\hyperlink{ref-boeing2019a}{Boeing, 2019a}). These metrics are
calculated for the entire street network and for each Superblock
individually. To facilitate further analysis, all of these metrics are
included in the exportable GeoPackage file.

\hypertarget{design}{%
\section{Design}\label{design}}

\texttt{superblockify}'s design is object-oriented with a focus on
modularity and extensibility. An abstract partitioner base class is
provided to facilitate implementing new custom approaches for Superblock
generation. At the core of the package, \texttt{superblockify} extends
Dijkstra's efficient distance calculation approach with Fibonacci heaps
on reduced graphs, ensuring optimal performance when iterating various
Superblock configurations while respecting the Superblock restriction of
no through traffic. This restriction is checked via just-in-time (JIT)
compilation through \texttt{numba}
(\protect\hyperlink{ref-siu_kwan_lam_2023_8087361}{Lam et al., 2023}) to
speed up the calculation of betweenness centrality on directed,
large-scale street networks. Central code dependencies are the
\texttt{osmnx} (\protect\hyperlink{ref-boeing2017}{Boeing, 2017}) and
\texttt{networkx} (\protect\hyperlink{ref-hagberg2008}{Hagberg et al.,
2008}) packages for data acquisition, preprocessing, and network
analysis, and the \texttt{geopandas}
(\protect\hyperlink{ref-joris_van_den_bossche_2023_8009629}{Bossche et
al., 2023}) package for spatial analysis.

\hypertarget{acknowledgements}{%
\section{Acknowledgements}\label{acknowledgements}}

Michael Szell acknowledges funding from the EU Horizon Project JUST
STREETS (Grant agreement ID: 101104240). All authors gratefully
acknowledge all open source libraries on which \texttt{superblockify}
builds, and the open source data that this software makes use of: Global
Human Settlement Layer, and map data copyrighted by OpenStreetMap
contributors available from \url{https://www.openstreetmap.org}.

\hypertarget{authors-contributions-with-credit}{%
\section{\texorpdfstring{Authors contributions with
\href{https://credit.niso.org/}{CRediT}}{Authors contributions with CRediT}}\label{authors-contributions-with-credit}}

\begin{itemize}
\tightlist
\item
  Carlson M. Büth: Conceptualization, Software, Investigation,
  Methodology, Writing -- original draft, Validation
\item
  Anastassia Vybornova: Conceptualization, Supervision, Writing --
  review \& editing, Validation
\item
  Michael Szell: Conceptualization, Project administration, Writing --
  review \& editing, Validation, Funding acquisition
\end{itemize}

\hypertarget{references}{%
\section*{References}\label{references}}
\addcontentsline{toc}{section}{References}

\hypertarget{refs}{}
\begin{CSLReferences}{1}{0}
\leavevmode\vadjust pre{\hypertarget{ref-agenciadecologiaurbanadebarcelona2021}{}}%
Agència d'Ecologia Urbana de Barcelona, Barcelona Regional Agència de
Desenvolupament Urbà, S.A., \& Àrea Metropolitana de Barcelona. (2021).
\emph{\href{http://hdl.handle.net/11703/122998}{BCNecologia: 20 años de
la Agencia de Ecología Urbana de Barcelona}}. {Ajuntament de Barcelona}.
ISBN:~978-84-9156-349-5

\leavevmode\vadjust pre{\hypertarget{ref-boeing2017}{}}%
Boeing, G. (2017). {OSMnx}: {New} methods for acquiring, constructing,
analyzing, and visualizing complex street networks. \emph{Computers,
Environment and Urban Systems}, \emph{65}, 126--139.
\url{https://doi.org/10.1016/j.compenvurbsys.2017.05.004}

\leavevmode\vadjust pre{\hypertarget{ref-boeing2019a}{}}%
Boeing, G. (2019a). The {Morphology} and {Circuity} of {Walkable} and
{Drivable Street Networks}. In L. D'Acci (Ed.), \emph{The {Mathematics}
of {Urban Morphology}} (pp. 271--287). {Springer International
Publishing}. \url{https://doi.org/10.1007/978-3-030-12381-9_12}

\leavevmode\vadjust pre{\hypertarget{ref-boeing2019}{}}%
Boeing, G. (2019b). Urban spatial order: Street network orientation,
configuration, and entropy. \emph{Applied Network Science}, \emph{4}(1),
67. \url{https://doi.org/10.1007/s41109-019-0189-1}

\leavevmode\vadjust pre{\hypertarget{ref-joris_van_den_bossche_2023_8009629}{}}%
Bossche, J. V. den, Jordahl, K., Fleischmann, M., McBride, J.,
Wasserman, J., Richards, M., Badaracco, A. G., Snow, A. D., Tratner, J.,
Gerard, J., Ward, B., Perry, M., Farmer, C., Hjelle, G. A., Taves, M.,
ter Hoeven, E., Cochran, M., Gillies, S., Caria, G., \ldots{} Ren, C.
(2023). \emph{Geopandas/geopandas: V0.13.2} (Version v0.13.2). {Zenodo}.
\url{https://doi.org/10.5281/zenodo.8009629}

\leavevmode\vadjust pre{\hypertarget{ref-brandes2008}{}}%
Brandes, U. (2008). On variants of shortest-path betweenness centrality
and their generic computation. \emph{Social Networks}, \emph{30}(2),
136--145. \url{https://doi.org/10.1016/j.socnet.2007.11.001}

\leavevmode\vadjust pre{\hypertarget{ref-buth2023master}{}}%
Büth, C. M. (2023). \emph{From gridlocks to greenways: {Analyzing} the
network effects of computationally generated low traffic neighborhoods}
{[}Master's thesis, University of M{ü}nster{]}.
\url{https://doi.org/10.13140/RG.2.2.26204.36481}

\leavevmode\vadjust pre{\hypertarget{ref-abstreet}{}}%
Carlino, D., Kirk, M., Smith, A., dcarlino, Konieczny, M., Kott, G.,
Bruce, Nissar, J., Nederlof, T., Steinberg, V., Lovelace, R., Ilias,
Nebeker, J., Sam, Orestis, Dejean, M., Shenfield, M., Schimek, N.,
Foucault, M., \ldots{} Huston, K. (2024). \emph{A-b-street/abstreet:
Night markets at -15C} (Version v0.3.49). Zenodo.
\url{https://doi.org/10.5281/zenodo.10476253}

\leavevmode\vadjust pre{\hypertarget{ref-eggimann_potential_2022}{}}%
Eggimann, S. (2022a). The potential of implementing superblocks for
multifunctional street use in cities. \emph{Nature Sustainability},
\emph{5}(5), 406--414. \url{https://doi.org/10.1038/s41893-022-00855-2}

\leavevmode\vadjust pre{\hypertarget{ref-eggimann_expanding_2022}{}}%
Eggimann, S. (2022b). Expanding urban green space with superblocks.
\emph{Land Use Policy}, \emph{117}, 106111.
\url{https://doi.org/10.1016/j.landusepol.2022.106111}

\leavevmode\vadjust pre{\hypertarget{ref-feng_algorithmic_2022}{}}%
Feng, C., \& Peponis, J. (2022). Algorithmic definitions of street
network centrality sub-shapes: {The} case of superblocks.
\emph{Environment and Planning B: Urban Analytics and City Science},
239980832210987. \url{https://doi.org/10.1177/23998083221098739}

\leavevmode\vadjust pre{\hypertarget{ref-frey2020potenziale}{}}%
Frey, H., Leth, U., \& Sandholzer, F. J. (2020). \emph{Potenziale von
superblock-konzepten als beitrag zur planung energieeffizienter
stadtquartiere-SUPERBE}.
\url{https://nachhaltigwirtschaften.at/en/sdz/projects/superbe.php}

\leavevmode\vadjust pre{\hypertarget{ref-hagberg2008}{}}%
Hagberg, A. A., Schult, D. A., \& Swart, P. J. (2008). Exploring
{Network Structure}, {Dynamics}, and {Function} using {NetworkX}. In G.
Varoquaux, T. Vaught, \& J. Millman (Eds.), \emph{Proceedings of the 7th
{Python} in {Science Conference}} (pp. 11--15).

\leavevmode\vadjust pre{\hypertarget{ref-kirkley2018}{}}%
Kirkley, A., Barbosa, H., Barthelemy, M., \& Ghoshal, G. (2018). From
the betweenness centrality in street networks to structural invariants
in random planar graphs. \emph{Nature Communications}, \emph{9}(1),
2501. \url{https://doi.org/10.1038/s41467-018-04978-z}

\leavevmode\vadjust pre{\hypertarget{ref-siu_kwan_lam_2023_8087361}{}}%
Lam, S. K., Pitrou, A., Florisson, M., Seibert, S., Markall, G.,
Anderson, T. A., Leobas, G., Collison, M., Bourque, J., Meurer, A.,
Oliphant, T. E., Riasanovsky, N., Wang, M., Pronovost, E., Totoni, E.,
Wieser, E., Seefeld, S., Grecco, H., Masella, A., \ldots{}
Turner-Trauring, I. (2023). \emph{Numba/numba: {Version} 0.57.1}
(Version 0.57.1). {Zenodo}. \url{https://doi.org/10.5281/zenodo.8087361}

\leavevmode\vadjust pre{\hypertarget{ref-latora2001}{}}%
Latora, V., \& Marchiori, M. (2001). Efficient {Behavior} of
{Small-World Networks}. \emph{Physical Review Letters}, \emph{87}(19),
198701. \url{https://doi.org/10.1103/PhysRevLett.87.198701}

\leavevmode\vadjust pre{\hypertarget{ref-laverty2021low}{}}%
Laverty, A. A., Goodman, A., \& Aldred, R. (2021). Low traffic
neighbourhoods and population health. In \emph{bmj} (Vol. 372). British
Medical Journal Publishing Group. \url{https://doi.org/10.1136/bmj.n443}

\leavevmode\vadjust pre{\hypertarget{ref-li_modeling_2023}{}}%
Li, K., \& Wilson, J. (2023). Modeling the {Health} {Benefits} of
{Superblocks} across the {City} of {Los} {Angeles}. \emph{Applied
Sciences}, \emph{13}(4), 2095. \url{https://doi.org/10.3390/app13042095}

\leavevmode\vadjust pre{\hypertarget{ref-martin2021}{}}%
Martin, R. J. (2021). \emph{Points of {Exchange}: {Spatial Strategies}
for the {Transition Towards Sustainable Urban Mobilities}}. {Aalborg
Universitetsforlag}. \url{https://doi.org/10.54337/aau451017237}

\leavevmode\vadjust pre{\hypertarget{ref-mattioli_political_2020}{}}%
Mattioli, G., Roberts, C., Steinberger, J. K., \& Brown, A. (2020). The
political economy of car dependence: {A} systems of provision approach.
\emph{Energy Research \& Social Science}, \emph{66}, 101486.
\url{https://doi.org/10.1016/j.erss.2020.101486}

\leavevmode\vadjust pre{\hypertarget{ref-miner2024car}{}}%
Miner, P., Smith, B. M., Jani, A., McNeill, G., \& Gathorne-Hardy, A.
(2024). Car harm: A global review of automobility's harm to people and
the environment. \emph{Journal of Transport Geography}, \emph{115},
103817. \url{https://doi.org/10.1016/j.jtrangeo.2024.103817}

\leavevmode\vadjust pre{\hypertarget{ref-mueller2020}{}}%
Mueller, N., Rojas-Rueda, D., Khreis, H., Cirach, M., Andrés, D.,
Ballester, J., Bartoll, X., Daher, C., Deluca, A., Echave, C., Milà, C.,
Márquez, S., Palou, J., Pérez, K., Tonne, C., Stevenson, M., Rueda, S.,
\& Nieuwenhuijsen, M. (2020). Changing the urban design of cities for
health: {The} superblock model. \emph{Environment International},
\emph{134}, 105132. \url{https://doi.org/10.1016/j.envint.2019.105132}

\leavevmode\vadjust pre{\hypertarget{ref-nieuwenhuijsen2019}{}}%
Nieuwenhuijsen, M., Bastiaanssen, J., Sersli, S., Waygood, E. O. D., \&
Khreis, H. (2019). Implementing {Car-Free Cities}: {Rationale},
{Requirements}, {Barriers} and {Facilitators}. In M. Nieuwenhuijsen \&
H. Khreis (Eds.), \emph{Integrating {Human Health} into {Urban} and
{Transport Planning}} (pp. 199--219). {Springer International
Publishing}. \url{https://doi.org/10.1007/978-3-319-74983-9_11}

\leavevmode\vadjust pre{\hypertarget{ref-nieuwenhuijsen_superblock_2024}{}}%
Nieuwenhuijsen, M., De Nazelle, A., Pradas, M. C., Daher, C., Dzhambov,
A. M., Echave, C., Gössling, S., Iungman, T., Khreis, H., Kirby, N.,
Khomenko, S., Leth, U., Lorenz, F., Matkovic, V., Müller, J., Palència,
L., Pereira Barboza, E., Pérez, K., Tatah, L., \ldots{} Mueller, N.
(2024). The {Superblock} model: {A} review of an innovative urban model
for sustainability, liveability, health and well-being.
\emph{Environmental Research}, \emph{251}, 118550.
\url{https://doi.org/10.1016/j.envres.2024.118550}

\leavevmode\vadjust pre{\hypertarget{ref-openstreetmapcontributors2023}{}}%
OpenStreetMap contributors. (2023). \emph{{OpenStreetMap}}.
\url{https://www.openstreetmap.org}

\leavevmode\vadjust pre{\hypertarget{ref-pesaresi2023}{}}%
Pesaresi, M., \& Politis, P. (2023). \emph{{GHS-BUILT-S R2023A} - {GHS}
built-up surface grid, derived from {Sentinel2} composite and {Landsat},
multitemporal (1975-2030)} {[}Data set{]}. {European Commission, Joint
Research Centre (JRC)}.
\url{https://doi.org/10.2905/9F06F36F-4B11-47EC-ABB0-4F8B7B1D72EA}

\leavevmode\vadjust pre{\hypertarget{ref-qgis}{}}%
QGIS Development Team. (2024). \emph{QGIS geographic information
system}. QGIS Association. \url{https://www.qgis.org}

\leavevmode\vadjust pre{\hypertarget{ref-ritchie2018}{}}%
Ritchie, H., \& Roser, M. (2018). Urbanization. \emph{Our World in
Data}. \url{https://ourworldindata.org/urbanization}

\leavevmode\vadjust pre{\hypertarget{ref-satterthwaite2009}{}}%
Satterthwaite, D. (2009). The implications of population growth and
urbanization for climate change. \emph{Environment and Urbanization},
\emph{21}(2), 545--567. \url{https://doi.org/10.1177/0956247809344361}

\leavevmode\vadjust pre{\hypertarget{ref-stadtwien2021}{}}%
Stadt Wien. (2021). \emph{Pilotstudie {Supergrätzl} - {Ergebnisbericht}
am {Beispiel Volkertviertel}} (GZ 367568; p. 58). {Stadt Wien -
Stadtentwicklung und Stadtplanung}.

\leavevmode\vadjust pre{\hypertarget{ref-szell2022}{}}%
Szell, M., Mimar, S., Perlman, T., Ghoshal, G., \& Sinatra, R. (2022).
Growing urban bicycle networks. \emph{Scientific Reports}, \emph{12}(1,
1), 6765. \url{https://doi.org/10.1038/s41598-022-10783-y}

\leavevmode\vadjust pre{\hypertarget{ref-transportforlondon2020}{}}%
Transport for London. (2020). \emph{Streetspace guidance: {Appendix Six}
(a): {Supplementary} guidance on {Low Traffic Neighbourhoods}}.
\url{https://tfl.gov.uk/info-for/boroughs-and-communities/streetspace-funding}

\leavevmode\vadjust pre{\hypertarget{ref-tuneourblock}{}}%
\emph{TuneOurBlock}. (2024).
\url{https://jpi-urbaneurope.eu/project/tuneourblock}.

\leavevmode\vadjust pre{\hypertarget{ref-world2022walking}{}}%
WHO. (2022). \emph{Walking and cycling: Latest evidence to support
policy-making and practice}.
\url{https://apps.who.int/iris/handle/10665/354589}

\leavevmode\vadjust pre{\hypertarget{ref-yan_redefining_2023}{}}%
Yan, X., \& Dennett, A. (2023). Re-defining {Transport} for {London}'s
strategic neighbourhoods from spatial and social perspectives.
\emph{Applied Geography}, \emph{161}, 103116.
\url{https://doi.org/10.1016/j.apgeog.2023.103116}

\end{CSLReferences}

\end{document}